\documentclass[10pt]{article}
\usepackage[centertags]{amsmath}
\usepackage{amsfonts}
\usepackage[all]{xy}
\usepackage{rotating}
\newtheorem{defi}{Definition}

\newtheorem{prop}{Proposition}
\newtheorem{cor}{Corollary}

\newtheorem{que}{Question}

\def\be{\begin{equation}}
\def\ee{\end{equation}}

\newcounter{examnum}[section]
\newcounter{remarnum}[section]
\begin{document}
\title{ Model Theory of Ultrafinitism I: \\ Fuzzy Initial Segments of Arithmetic\\ (Preliminary Draft)}
\author{Mirco A. Mannucci   \\
Rose M. Cherubin }

\date{\today}
\maketitle
\begin{abstract}
\noindent
 This article is the first of an intended series of works
on the model theory of Ultrafinitism. It is roughly divided into two
parts. The first one addresses some of the issues related to
ultrafinitistic programs, as well as some of the core ideas proposed
thus far. The second part of the paper  presents a model of
ultrafinitistic arithmetics based on the notion of fuzzy initial
segments of the standard natural numbers series. We also introduce a
proof theory and a semantics for ultrafinitism through which
feasibly consistent theories can be treated on the same footing as
their classically consistent counterparts. We conclude with a brief
sketch of a foundational program, that aims at reproducing the
transfinite within the finite realm.

\end{abstract}
\section{Preamble}
\begin{quote}
    To the memory of our unforgettable friend Stanley "Stan"  Tennenbaum
    ($1927-2005$), Mathematician, Educator, Free Spirit
\end{quote}

As we have mentioned in the abstract, this article is the first one
of a series dedicated to ultrafinitistic themes. \\
\\
First papers often tend to take on the dress of manifestos, road
maps, or both,  and this one is no exception. It is the revised
version of an invited  conference talk, and was meant from the start
for a quite large audience of philosophers, logicians, computer
scientists, and mathematicians, who might have some interest in the
ultrafinite. Therefore, neither the philosophico-historical, nor the
mathematical side, are meant to be detailed investigations. Instead,
a number of items, proposals, questions, etc. are raised, which will
be further explored  in subsequent works of the series.
\\\\
Our chief hope is that readers will find the overall "flavor"  a bit
"Tennenbaumian". And friends of Stan, old and new,  know what we
mean \dots

\section{Introduction: the Radical Wing of Constructivism}

In their encyclopedic work on Constructivism in Mathematics
(\cite{Troelstra}), A. Troelstra and D. Van Dalen dedicate only a
small section to Ultrafinitism (UF in the following). This is no
accident: as they themselves explain therein, there is no consistent
model theory for ultrafinitistic mathematics. It is well-known that
there is a plethora of models for intuitionist logic and mathematics
(realizability models, Kripke models and their generalization based
on category theory, etc.). Thus, a skeptical mathematician who does
not feel committed to embrace the intuitionist faith (and most do
not), can still understand and enjoy the intuitionist's viewpoint
while remaining all along within the  confines of classical
mathematics. Model theory creates, as it were, the bridge between
quite different worlds.
\\
\\
It would be desirable that something similar were available for the
more radical positions, that go under the common banner of
Ultrafinitism. To be sure, in the fifteen years since the
publication of the cited book, some proposals have emerged to fill
the void. We still feel, though, that nothing comparable to the
sturdy structure of  model theory for intuitionism is available thus
far. This article is the first one in a series that aims at
proposing several independent but related frameworks for UF.
\\
\\
Before embarking on this task, though, an obvious question has to be
addressed first:
\\
\\
What is Ultrafinitism, really?
\\
\\
It turns out that to provide a satisfactory answer is no trivial
task. Of course, one could simply answer: all positions in
foundation of mathematics that are more radical than traditional
constructivism (in its various flavors). This answer would beg the
question: indeed,  {\it what} makes a foundational program that
radical?
\\
\\
There is indeed at least one common denominator for ultrafinitists,
namely the deep-seated \emph{mistrust for all kinds of infinite,
actual and potential alike}. Having said that, it would be tempting
to conclude that UF is quite simply the rejection of infinity in
favor of the study of finite structures (finite sets, finite
categories), a program that has been partly carried out in some
quarters. For instance in Finite Model Theory (see the excellent
online notes written by Jouko Vannaenen \cite{Jouko}).
\\
\\
Luckily (or unluckily, depending on reader's taste), things are not
that straightforward, on account of at least two substantive points:
\begin{itemize}
  \item First, the rejection of infinitary methods, even the ones based on
  the so-called potential infinite, must be applied at {\it all levels}, including
  meta-mathematics and logical rules. Both syntax and semantics must
  be changed to fit the ultrafinitistic paradigm. Approaches such as
  Finite Model Theory are simply not radical enough for the task a
  hand, as they are still grounded in a semantics and syntax that is
  deeply entrenched with the infinite.

  \item Second, barring one term in the dichotomy finite-infinite,
  is, paradoxically, an admission of guilt: the denier implicitly
  agrees that the dichotomy itself stands solid. But does it? Perhaps the
  black-white chessboard should be replaced with various tonalities
  of grey.

\end{itemize}
These two points must be addressed by a convincing model theory of
Ultrafinitism. This means that such a model theory (assuming that
anything like it can be produced), must be able to generate
classical (or intuitionistic)  structures, let us call them
\emph{ultrafinitistic universes}, wherein an hypothetical
ultrafinitist mathematician can live forever happily. The dweller of
those universes should be able to treat certain finite objects as
\emph{de facto} infinite.
\\
\\
Logicians are quite used to the-- \emph{inside versus outside}--
pattern of thought: it suffices to think of the minimal model of
$ZF$ to get the taste (inside the minimal model countable ordinals
look and feel like enormous cardinals, to quote just one blatant
effect). It seems that, if we could "squeeze" the minimal model
below $\aleph_0$, we could get what we are looking for.
\\
\\
Only one major obstacles stands in the way: \emph{the apparently
absolute
character of the natural number series}. \\
\\
But it is now time for a bit of history \ldots

\section{Short History and Prehistory of Ultrafinitism.}
\begin{quote}
    \emph{The trouble with eternity is that one never knows when it will
    end.}\\
        Sam Stoppard, Rosenkranz and Guildenstern Are Dead.
\end{quote}
Ultrafinitism has, quite ironically, a very long prehistory. Indeed,
it extends into and encroaches the domains of cultural anthropology
and child cognitive psychology: some "primitive" cultures, and
children alike, do not seem to have even a notion of arbitrarily
large numbers. To them, the natural number series looks a bit like:
-One, two, three, \ldots , many!-. An exploration of these alluring
territories would bring us too far afield, so we shall restrict our
tale to the traditional beginning of western culture, the Greeks
(our version here is only a sketch. For a more exhaustive and
analytical investigation, see our \cite{Cherubin}).
\\
\\
Ancient Greek mathematical work does not explicitly treat the
ultrafinite. It is therefore all the more interesting to note that
early Greek poetry, philosophy, and historical writing incorporate
two notions that are quite relevant for the study of the
 ultrafinite. These notions are epitomized by two words:
 \textbf{Murios } ($\mu\upsilon\rho\iota o\varsigma$) and  \textbf{ Apeir\={o}n}
 ($\alpha\pi\epsilon\iota\rho\omega\nu$).
\\
\\
\textbf{Murios}
\\
\\
The term murios has two basic senses, each of which is used in
specific ways. These senses are 'very many'/ 'a lot of';  and 'ten
thousand.' The first sense denotes an aggregate or quantity whose
exact number is either not known or not relevant; the second denotes
a precise number. With some exceptions, to be detailed below, the
syntax and context make clear which sense is intended in each case.
It is part of the aim of this paper to draw attention to the
importance of contextualized usage in understanding the ultrafinite.
\\
\\
The earliest occurrences of the term \emph{murios}  appear in the
oldest extant Greek writing, viz., Homer's Iliad and Odyssey. In
Homer, all 32 instances of forms of \emph{murios}  have the sense
'very many'/ 'a lot of.' Translations often render the word as
'numberless,' 'countless,' or 'without measure.' What exactly does
this mean: does \emph{murios}  refer to an indefinite number or
quantity, to an infinite number or quantity, to a number or quantity
that is finite and well-defined but that is not feasibly countable
for some reason, or to a number or quantity that the speaker deems
large but unnecessary to count? Our investigation reveals that Homer
tends to use the term in the last two ways, that is, to refer to
numbers or quantities for which a count or measure would be
unfeasible, unnecessary, or not to the point. In general, Homer uses
the word in situations where it is not important to know the exact
number of things in a large group, or the exact quantity of some
large mass.
\\
\\
Some representative examples of Homer's usages of \emph{murios} :
(a) At Iliad 2.468, the Achaeans who take up a position on the banks
of the Scamander are murioi (plural adjective), "such as grow the
leaves and flowers in season." The leaves and flowers are certainly
not infinite in number, nor are they indefinite in number, but they
are not practicably countable, and there is no reason to do so - it
is enough to know that there are very many all over. (b) In the
previous example, the Achaeans must have numbered at least in the
thousands. \emph{murios}  can, however, be used to refer to much
smaller groups. At Iliad 4.434, the clamoring noise made by the
Trojans is compared to the noise made by muriai ewes who are being
milked in the courtyard of a very wealthy man (the ewes are bleating
for their lambs). The number of ewes owned by a man of much property
would certainly be many more than the number owned by someone of
more moderate means, but that rich man's ewes - especially if they
all fit in a courtyard - must number at most in the low hundreds.
This suggests that the ewes are said to be muriai in number because
there are a comparatively large number of them; because there is no
need to count them (a man who had 120 ewes and was considered very
wealthy would not cease to be considered very wealthy if he lost one
or even ten of them; the Sorites paradox is later); and possibly
because it might not be practicable to count them (they might be
moving around, and they all look rather like one another).
\\
\\
Similarly, at Odyssey 17.422 Odysseus says he had murioi slaves at
his home in Ithaca before he left for the Trojan war. The word for
'slaves' in this case is \emph{dm\={o}es}, indicating that these are
prisoners of war. Given what we know of archaic Greek social and
economic structures, the number of slaves of this type that a man in
his position could have held must have been in the dozens at most.
The key to Odysseus' use of the term is the context. The sentence as
a whole reads, "And I had murioi slaves indeed, and the many other
things through which one lives well and is called wealthy." That is,
the quantities of slaves and of other resources that he commanded
were large enough to enable him to be considered wealthy. The exact
number of slaves might have been countable, but it would have been
beside the point to count them. (c) \emph{murios}  can also refer to
quantities that are not such as to be counted. At Odyssey 15.452, a
kidnapped son of a king is projected to fetch a \emph{murios}  price
as a slave. Here \emph{murios} must mean 'very large,' 'vast.' This
is by no means to say that the price will be infinite or indefinite,
for a price could not be thus. Rather, the situation is that the
exact price cannot yet be estimated, and the characters have no need
to estimate it (i.e., they are not trying to raise a specific amount
of money). Prices are not countable, but they are of course
measurable or calculable. There are also instances of \emph{murios}
in Homer that refer to kinds of things that do not seem to be
measurable or calculable. At Iliad 18.88, Achilles says that his
mother Thetis will suffer \emph{murios}  grief (penthos) at the
death of Achilles, which is imminent. At 20.282, \emph{murios}
distress comes over the eyes of Aeneas as he battles Achilles. \\
\\
The usual translation of \emph{murios}  here is 'measureless'. This
translation may be somewhat misleading if it is take literally, as
there is no evidence that the Greeks thought that smaller amounts of
grief and distress were necessarily such as to be measurable or
measured. A more appropriate translation might be 'vast' or
'overwhelming.' It is possible that Achilles means that Thetis will
suffer grief so vast that she will never exhaust it nor plumb its
depths even though she is immortal; but it is also possible that
Homer did not consider whether grief or distress could be unending
and infinite or indefinite in scope. \\
\\
The epic poet Hesiod (8th-7th BCE) and the historian Herodotus (5th
BCE) sometimes use \emph{murios}  in the senses in which Homer does,
but they also use it to mean ten thousand. With a very few
exceptions, the syntax and context make clear in each instance which
meaning is present. At Works and Days 252, Hesiod says that Zeus has
tris murioi immortals (i.e. divinities of various kinds) who keep
watch over mortals, marking the crooked and unjust humans for
punishment. \emph{Tris} means three times or thrice, and there is no
parallel in Greek for understanding \emph{tris murioi} as "three
times many" or "three times a lot." There are parallels for
understanding \emph{tris} with an expression of quantity as three
times a specific number; and the specific number associated with
\emph{murios}  is ten thousand. Therefore \emph{tris} \emph{murios}
should indicate thirty thousand. Some instances of \emph{murios}  in
Herodotus clearly refer to quantities of ten thousand; some clearly
refer to large amounts whose exact quantities are unspecified; and a
few are
ambiguous but do not suggest any meaning other than these two. \\
\\
(d) At 1.192.3, Herodotus says that the satrap Tritantaechmes had so
much income from his subjects that he was able to maintain not only
warhorses but eight hundred (\emph{oktakosioi}) other breeding
stallions and \emph{hexakischiliai kai muriai mares}.
\emph{Hexakischiliai} means six times one thousand, so that the
whole expression should read six thousand plus \emph{muriai}. The
next line tells us that there are twenty (\emph{eikosi}) mares for
every stallion, so that the total number of mares must be sixteen
thousand, and \emph{muriai} must mean ten thousand (it is a plural
adjective to agree with the noun). The case is similar at 2.142.2-3.
Here Herodotus says that three hundred (\emph{tri\={e}kosiai})
generations of men come to muriai years since three generations come
to one hundred (\emph{hekaton}) years. Clearly, \emph{muriai} means
ten thousand here.
\\
\\
(e) In some places, Herodotus cannot be using \emph{murios} to mean
ten thousand, and it is the context that shows this. At 2.37.3, for
example, describing the activities of Egyptian priests, he says that
they fulfill \emph{muriai} religious rituals, \emph{h\={o}s eipein
log\={o}i}. He may in fact mean that they fulfill \emph{muriai}
rituals each day, since the rest of the sentence speaks of their
daily bathing routines. Herodotus does not give any details about
the rituals or their number, and \emph{h\={o}s eipein log\={o}i}
means "so to speak." Thus Herodotus seems to be signalling that he
is not giving an exact figure, and \emph{muriai} must simply mean "a
great many." At 2.148.6, Herodotus reports that the upper chambers
of the Egyptian Labyrinth \emph{th\={o}ma murion pareichonto},
furnished much wonder, so remarkably were they built and decorated.
Certainly no particular amount of wonder is being specified here.
\\
\\
(f) Some occurrences of \emph{murios}  are ambiguous in a way that
is of interest for the study of the ultrafinite. At 1.126.5, Cyrus
sets the Persians the enormous task of clearing an area of eighteen
or twenty stadia (2 1/4  or 2 1/2 miles) on each side in one day,
and orders a feast for them the next. He tells them that if they
obey him, they will have feasts and muria other good things without
toil or slavery, but that if they do not obey him, they will have
\emph{anarithm\={e}toi} toils like that of the previous day. That
is, Cyrus is contrasting muria good things with
\emph{anarithm\={e}toi} bad ones. Is he asking the Persians to
consider this a choice between comparable large quantities? If so, a
\emph{murios} amount would be anarithm?tos, which can mean either
"unnumbered" or "innumerable, numberless." It is also possible that
\emph{murios} is supposed to mean ten thousand, so that the
magnitude of the undesirable consequences of defying Cyrus is
greater than the great magnitude of the advantages of obeying him.
If that is the meaning, Herodotus may be using \emph{murios}  in a
somewhat figurative sense, as when one says today that one has "ten
thousand things to do today." \emph{murios} , then, referred in the
earliest recorded Greek thought to large numbers or amounts. When it
did not refer to an exact figure of ten thousand, it referred to
numbers or amounts for which the speaker did not have an exact count
or measurement. Our analysis indicates that the speaker might lack
such a count or measurement either because the mass or aggregate in
question could not practicably be counted or measured under the
circumstances, or because an exact count or measurement would not
add anything to the point the speaker was making. In most cases it
is clear that the numbers and amounts referred to as \emph{murios}
were determinate and finite, and could with appropriate technology
be counted or measured. In instances where it is not clear whether
that which is referred to as \emph{murios}  is supposed to be such
as to admit of measuring or counting (Thetis' grief, for example;
and Cyrus' \emph{murios} good things if they are comparable to the
\emph{anarithm\={e}tos}), there is no evidence as to whether the
\emph{murios} thing or things are supposed to be infinite or
indefinite in scope. Indeed, there is no evidence that these early
writers thought about this point. (This is perhaps why
\emph{anarithm\={e}tos} can mean both "unnumbered" and
"innumerable," and why it is often difficult to tell which might be
meant and whether a writer has in mind any
distinction between them.) \\
\\
When \emph{murios}  does not mean "ten thousand," context determines
the order of quantity to which it refers. Any number or amount that
is considered to be "a lot" or "many" with respect to the
circumstances in which it is found can be called \emph{murios} .
Leaves and flowers in summer near the Scamander number many more
than those of other seasons, perhaps in the millions; but the rich
man's ewes are \emph{muriai} too, even if they number perhaps a
hundred. They are several times more than the average farmer has,
and they may fill the courtyard so much that they cannot easily be
counted.
    \\
    \\
\textbf{Apeiron}
\\
\\
Since \emph{murios}  seems to refer overwhelmingly to determinate
and finite quantities, it is useful to note that Greek had ways of
referring to quantities that were indeterminate, unlimited,
indefinite, or infinite. The most significant of these, for
philosophical and mathematical purposes, was the word apeiros or
apeir\={o}n (m., f.)/ apeiron (n.). \\
\\
The etymology of this word is generally understood to be
\emph{peirar} or \emph{peras}, "limit" or "boundary," plus
alpha-privative, signifying negation: literally, "not limited" or
"lacking boundary." Etymology alone does not tell us the range of
uses of the term or the ways in which it was understood, so we must
again consider its occurrences in the earliest sources. The term
appears as early as Homer, in whose poems it generally refers to
things that are vast in extent, depth, or intensity.
\\
\\
Homer uses \emph{apeiros} most frequently of expanses of land or
sea. In each case, the apeiros/apeir\={o}n  thing is vast in breadth
or depth; whether its limits are determinable is not clear from the
context, but limits do seem to be implied in these cases. Some
instances may imply a surpassing of some sort of boundaries or
borders (though not necessarily of all boundaries or borders). At
Iliad 24.342 and Odyssey 1.98 and 5.46, a god swiftly crosses the
apeiron earth. Within the context, it is clear that the poet means
that the divinity covers a vast distance quickly. There may be a
further implication that the gods transcend or traverse boundaries
(be these natural features or human institutions) with ease, so that
the world has no internal borders for them. Similarly, in Od.
17.418, the expression \emph{kat' apeirona gaian}, often translated
as "through[out] the boundless earth," is used to suggest that
something is spread over the whole earth. What is spread covers a
vast expanse, and it also crosses all boundaries on the earth. Two
other Homeric examples are of interest. At Od. 7.286, a sleep is
described as apeiron, meaning either that it is very deep, or
unbroken, or both. At Od. 8.340, strong bonds are apeiron,
surpassing limits of a god's
strength, and so unbreakable. \\
\\
Hesiod also uses apeiron to describe things that extend all over the
earth, but also uses the word once in reference to a number. In
Shield of Heracles 472, the word refers to a large number of people
from a great city involved in the funeral of a leader; the sense
seems to be that there were uncountably many, and possibly that the
leader's dominion had been vast.
\\
\\
Herodotus (5th century BCE) uses apeiron in two cases where its
meaning clearly derives from  the privative of peirar.
\footnote{Herodotus also uses a homophone word that is derived from
another root, so we have only included instances where context
clearly indicates that the word is the one derived from peirar.  }
In 5.9 he uses it to refer to a wilderness beyond Thracian
settlements. In 1.204 a plain is apeiron, perhaps hugely or
indeterminately vast. In both cases, Herodotus knows that the lands
are definite in extent (he identifies the peoples who live beyond
them). The contexts suggest that he means that these lands are vast
and that their exact boundaries are not known. He may also have in
mind that they cannot be easily, if at all, traversed by humans.
\footnote{Kahn (\cite{Kahn}) argues that "[t]he true sense of
apeiros is therefore 'what cannot be passed over or traversed from
end to end.'" He admits, however, that Homer's gods traverse the
earth and sea that Homer calls apeiron. For the term to have the
sense of 'untraversible,' it would have to be understood as an
exaggeration in Homer and Herodotus. Gods and rivers, for example,
do traverse the things described as apeiron, but it is extremely
difficult or impractical for humans to do so. An alternative to this
interpretation would be to say that the term does not denote
untraversibility, though it may connote that in some cases, and that
it instead denotes vastness and often practical indeterminability.}
\\
\\
The first and perhaps best-known philosophical use of apeiron is in
the reports about the work of Anaximander in the sixth century BCE.
Anaximander is reported to have held that the source of all familiar
things, the fundamental generative stuff of the cosmos, was
something apeiron. The testimonia report that the apeiron was
eternal in duration, unlimited or indeterminate in extent, and
qualitatively indeterminate.\footnote{Anaximander's fundamental
entity is generally referred to as 'the apeiron,' since he did not
give it another name and did not say what it was in any other way.
Adjectives, especially neuter adjectives, are readily used as
substantives in ancient Greek.    }  It was neither fire, nor water,
nor air, nor earth, nor hot, cold, light, dark, etc.; but it was
that which could give rise to all of these. \footnote{ Kahn
(\cite{Kahn}) holds that Anaximander's apeiron is "primarily a huge,
inexhaustible mass, stretching away endlessly in every direction."
It must at least be that, but as Richard McKirahan points out in
Philosophy Before Socrates, the discussions of Anaximander in
Aristotle and the Peripatetics make clear that the apeiron must also
be a stuff of indefinite kind or quality (33-35).} All of the
familiar cosmos, for Anaximander, arose from the apeiron.
\\
\\
We may note that so far no instance of apeiron clearly means
'infinite.' Only one, Anaximander's, could possibly involve an
infinite extent, and even in that case it is not clear that the
extent is infinite; it may be indefinite or inexhaustible without
being infinite. Anaximander's stuff is eternal, i.e. always in
existence, but it is not at all clear that a sixth century Greek
would have taken "always" to mean an infinite amount of time.
Whether any Greek of the 8th to 5th centuries BCE conceived of
quantities or magnitudes in a way that denoted what we would call
infinity is not certain.\\
\\
It is sometimes thought that Zeno of Elea (5th century) spoke of the
infinite, but there is good evidence that he had quite a different
focus. It is only in the arguments concerning plurality that are
preserved by Simplicius that we find what may be quotations from
Zeno's work (regarding his arguments concerning motion and place we
have only reports and paraphrases or interpretations).\footnote{  It
is possible that some of Zeno's paradoxes of motion dealt with
infinitely long sequences of steps. Aristotle suggests that they
did. Aristotle used the word apeiron to describe these sequences,
but it is not known whether Zeno did. See Aristotle, Physics Z2
\cite{Ari}.} In fragments DK29 B1 and B2, Zeno argued "from saying
that multiple (polla, many) things are, saying opposite things
follows." In particular, if we say that multiple things are, then we
must conclude that "the same things must be so large as to be apeira
(neuter plural) and so small as to lack magnitude (megethos)." Zeno
was evidently interested in the claim that there are multiple things
with spatial magnitude, and it appears from the fragments that he
thought that the possibilities for analyzing the components of
spatial magnitude were the hypothesis that a thing that has spatial
magnitude must be composed of parts with positive spatial magnitude,
parts of no magnitude, or some combination of these. If a thing had
no magnitude, Zeno argued, it would not increase (in magnitude)
anything to which it was added, nor decrease anything from which it
was removed. Therefore it could not "be" at all (at least, it could
not "be" as the spatial thing it was said to be). Nothing with
magnitude could be composed entirely of such things. However, if we
assume that the components of a spatial thing have positive
magnitude, another problem arises. In measuring such a thing, we
would try to ascertain the end of its "projecting part" (i.e. the
outermost part of the thing). Each such projecting part would always
have its own projecting part, so that the thing would have no
ultimate "extreme" (eschaton). That is, the outer edge of something
always has some thickness, as do the lines on any ruler we might use
to measure it; and this thickness itself can always be divided. Thus
the magnitude of a spatial thing, and thus its exact limits, will
not be determinable. There is nothing in this to suggest that Zeno
thought that the claim that there are multiple spatial things led to
the conclusion that such things must be infinitely large. Rather,
his description suggests that the things would be indeterminable,
and indeterminate or indefinite, in size.\footnote{For an account of
what Zeno showed, and what he may have been trying to show, in these
arguments, see \cite{Cherubin}, pages 5-7.  }
\\
\\
In the philosophy of the fourth century BCE, and arguably as early
as Zeno, an apeir\={o}n quantity could not be calculated exactly, at
least as long as it was regarded from the perspective according to
which it was apeir\={o}n. In fact, Aristotle's argument that a
continuous magnitude bounded at both ends could be traversed in a
finite amount of time - despite the fact that it "contains" an
apeir\={o}n number of points, and despite Zeno's Dichotomy argument
- rests precisely on the notions that the magnitude is not composed
of the apeir\={o}n number of points, and that from one perspective
it is bounded. Aristotle does not refute Zeno's argument, but merely
argues that within the framework of his physics, the question Zeno
addresses can be put into different terms. Thus where \emph{murios}
did not clearly refer to ten thousand, a \emph{murios}  quantity was
generally recognized as definite but was not calculated exactly. An
apeir\={o}n thing or quantity in Homer or Herodotus might be
definite or not, and in later thinkers, especially in philosophy,
the term came to emphasize that aspect of the thing or group or
quantity that was indefinite, indeterminate, or unlimited.
\\
\\
\textbf{Intermezzo}
\\
\\
We shall see momentarily that subsequent reflections on the
ultrafinite orbit, for the most part, around the myrios-apeiron
pair, as if around a double star. Meanwhile, to keep the length of
this paper a feasible one, we must skip over two thousand years of
mathematical and philosophical thought (where ultrafinitistic themes
crop up from time to time in the philosophical and mathematical
debate), and pick up the thread once again well into the twentieth
century.
\\
\\
\textbf{Recent History of UF}
\\
\\
The passage from prehistory to the history of UF is difficult to pin
down. Perhaps a bit arbitrarily, we shall say that it starts with
the radical criticism of Brouwer's Intutionist Programme  by Van
Dantzig \cite{VanDantzig} in 1950. \footnote{Van Dantzig himself
points out that some of his ideas were anticipated by the Dutch
philosopher Mannoury, and by the French mathematician Emil Borel
(\cite{Borel}). Borel observed that large finite numbers (\emph{les
nombres inaccessibles}) present the same order of difficulties as
the infinite. } This seminal  small paper contains \emph{in nuce}
most of the later motives. Chief among them that (quoting him
verbatim):\\
\\
- \emph{the difference between finite and infinite
numbers is not an essential, but a gradual one} -.
\\
\\
According to this view, an infinite number is a number that
surpasses everything I can ever obtain. One is here reminded of a
game inadvertently initiated by the Greek mathematician Archimedes
in his Sand Reckoner ($\Psi\alpha\mu\mu\iota\tau\varepsilon\varsigma
$. A good translation is \cite{Archimedes}), a game that is still
played nowadays: two fellows, A and B, trying to beat each other at
naming huge numbers. The winner is the contestant who is able to
construct a faster growing primitive recursive function. The winner
is the (temporary) owner of "infinite numbers" (see on this
\cite{BigNumbers}).
\\
\\
A second major character in this story is the Russian logician
Yessenin-Volpin. In a series of papers (see for instance his 1970
manifesto \cite{Volpin} )he exposes his views on UF. Unfortunately,
in spite of their appeal, his views are difficult to articulate:
there is simply too much there (though, to be sure, David Isles has
made a serious and quite successful attempt to clarify some of
Volpin's tenets in \cite{Isles1}). One of the fundamental ideas put
forth by Volpin is that \emph{there is no uniquely defined natural
number series}. Volpin's ruthless attach unmasks the circularity
behind the induction scheme, and leaves us with various not
isomorphic finite natural number series. \\
\\
Interestingly, a few years later, Princeton mathematician Ed Nelson
had an epiphany, described in his "confession" \cite{Nelson}: in a
morning of the Fall 1976, in Canada, Nelson lost his "pythagorean
faith" in the natural numbers. What was left was nothing more than
finite arithmetic terms, and the rules to manipulate them. Nelson's
Predicative Arithmetics (\cite{Nelson2}), albeit an essential steps
toward the re-thinking of mathematics along strictly finitist lines,
seems to us not as radical as Nelson's amazing vision: why stopping
at induction over bounded formulae? If numbers are no more, and
arithmetics is a concrete manipulation of symbols (a position that
could be aptly called ultra-formalist), "models" of arithmetics are
conceivable, where even the successor operation is not total, and
all induction is either restricted or banished altogether(for an
analysis of Nelson's arithmetics and for our proposal of an
arithmetics for ultra-formalists  see \cite{Mannucci3}).
\\
\\
Enter Parikh \cite{Parikh}. His celebrated "Existence and
Feasibility in Arithmetics" (1971) introduces an expanded version of
Peano Arithmetic,  enriched with a unary predicate F (where F(x)
intended meaning is that x is a feasible number, in some unspecified
sense). Mathematical induction does not apply to F, and a new axiom
is added to PA saying that a very large number is not feasible. More
precisely, the axiom says that the number $2_{1000}$, where $2_0 =
1$ and $2_{k +1} = 2^{2_k}$, is not feasible. Parikh proves, among
other important results, that the theory $PA + \neg F(2_{1000})$ is
\emph{feasibly consistent}: though obviously inconsistent from the
classical standpoint, all proofs exhibiting its inconsistency are
unfeasible, in the sense that the length of any such proof is a
number $n\geq 2_{1000}$.
\\\\
From the point of view of this history, Parikh achieves at least two
goals: first, he turns some of the claims of ultrafinitists into
concrete verifiable theorems. Secondly, he paves the way to a new
kind of \emph{ultrafinitistic proof theory}. Parikh's approach has
been improved by several authors (\cite{Dragalin},
\cite{Gavrilenko}). Quite recently, Vladimir Sazonov \cite{Sazonov},
has made a serious attempt at making more explicit the structure of
Ultrafinitistic Proof Theory.\footnote{In the cited paper the
absolute character of being a feasible number is asserted, on some
physical ground (Sazonov's actual position on this issue is more
articulate, as it appears from his recent FOM postings). We do not
share this belief. As pointed out elsewhere in this paper, we think
it is important to maintain the notion of contextual feasibility
(after all, who knows for sure what is out there? Perhaps new
discoveries in Plank level physics will show that the estimated
upper bound of "particles" in the universe was way too small.
Whereas logic should be able to account for physical limitations, it
should not be enslaved by them.)}. Also, Alessandra Carbone  and
Stephen Semmes, have investigated the consistency of $PA + \neg
F(2_{1000})$ and similar theories from a novel proof theoretical
standpoint, involving the combinatorial complexity of proofs
(\cite{Carbone}). We shall touch upon these ideas in Section 5.
\\
\\
Parikh's seminal paper leaves us with a desire for more: knowing
that $PA + \neg F(2_{1000})$ is feasibly consistent, there ought to
be some way of saying that it has a model. In other words, the
suspicion rises that, were a genuine semantics of ultrafinitistic
theory available to us, \emph{the celebrated G\H{o}del's
completeness theorem (or a finitist version thereof) should hold
true}. But where to look for such a semantics? Models are structured
sets (or objects in some category with structure). We must thus turn
our sight from proof theory to set theory and category theory.
\\
\\
On the set-theoretical side, there are at least two major
contributions. The first one is Vopenka's programme to reform
Cantor's Set Theory, also known as Alternative Set Theory (AST). AST
has been developed for more than three decades, so even a scanty
exposition of his results is not feasible here. We can just recall
the main themes: AST is a phenomenological theory of finite sets.
Some sets can have subclasses that are not themselves sets. Sets of
that kind are "infinite" in Vopenka's sense. Here we go back to one
of the senses of the word apeiron, previously described: some sets
are (or appear) infinite because they live outside our perceptual
horizon. It should be pointed out that AST is not, per se, a UF
framework. However, Vopenka envisioned the possibility of "witnessed
universes", i.e. universes where infinite (in his sense) semisets
contained in finite sets do exist. These witnessed universes would
turn AST into a "universe of discourse" for ultrafinitism. To our
knowledge, though, witnessed AST has not been developed beyond its
initial stage.
\\
\\
There are other variants of set theory with some finitist flavor.
For instance, Andreev and Gordon \cite{AndreevGordon} describe a
theory oh Hyper-finite Sets (THS), which is not incompatible with
classical set theory. An interesting fact is that both AST and THS
produce as a by-product a quite natural model of non-standard
analysis, \footnote{The very large and the very small are indeed
intimately related: if one has a consistent notion of large,
unfeasible number $n$, one automatically gets the "infinitesimal"
$\frac{1}{n}$, via the usual construction of the field of fractions
$\mathcal{Q}$. } thereby providing at least a good reason for
interest in finitistic math for mainstream mathematicians.
\\
\\
The second set-theoretical approach we are aware of, is the one
described in the book of Shaughan Lavine \cite{Lavine}. Here, a
finitistic variant of ZF is introduced, where the existence of a
large number, the Zillion, is posited (again, the reader may recall,
a old motive: the coming back of myrion.  Zillion replaces in some
way the missing $\aleph_{0}$)
\\
\\
Finally, category theory. This is, with one notable exception, a
totally uncharted (but very promising) territory. The exception is a
couple of works in the late seventies and early eighties by the late
Jon M. Beck, on using simplicial and homotopic methods to model
finite, concrete analysis \cite{Beck1}. As far as we understand it,
Beck's core idea is to use the simplicial category $\triangle$,
truncated at a certain level $\triangle[n]$, to replace the role of
the natural number series (or, because we are here in a categorical
framework, the so-called natural number object that several topoi
possess). The truncated simplicial category has enough structure to
carry out some finitistic version of recursion; moreover, its
homotopy theory provides new tools to model finite flow diagrams.
\footnote{ As it has been pointed out in a recent posting by Michael
Barr on the category list in appreciation of Beck's mathematics,
addition for the finite calculator is not an associative operation.
Homotopy "repairs" the lack of associativity by providing
"associativity up to homotopy" via coherence rules.} Beck's approach
will be discussed in a later work \cite{Mannucci3}, together with
other possible paths using category theory and the topos approach to
realizability in particular. \footnote{ As it is well known, topoi
have an internal logic, which is intuitionistic. One can hope that,
by isolating "feasible" objects in the realizability topos via a
suitable definition of \emph{feasible realizability}, a categorical
universe of discourse for UF could be, as it were, carved out.}
\\
\\
Before moving on to our first sketch of a proposal, let us try to
sum up some lessons we believe can be learned from the foregoing:

\begin{itemize}

  \item First, the notion of feasibility should be
  \emph{contextual}. An object (a term, a number, a set) is feasible
  only within a specific context, that specifies the type of resources
  available (functions, memory, time, etc.). Thus a full-blown model
  theory of UF should provide the framework for a dynamic notion of
  feasibility.
  \item As the contest changes, so does feasibility. What was unfeasible
  before, may become feasible now. Perhaps our notion of potential
  infinity came as the realization (or faith) that \emph{any} contest can be
  transcended.
\item The \emph{transcendency degrees of feasibility} are not necessarily
linearly ordered. One can envision contests in which what is
feasible for A is not feasible for B, and viceversa.
\item Lat, but not least, the pair Murios-Apeiron. Every convincing
approach to UF should be broad enough to encompass both terms. Even
better, it should unify the two streams into a single, flexible
framework.
\end{itemize}

\section{Fuzzy Initial Segments of NN: a model  of
ultrafinitistic arithmetics} In our first model we shall adopt a
moderate position, namely we still assume that "numbers" exist, at
least in the background (a more radical step, where the term model
of arithmetic is de-constructed into its partial fragments, and
where  numbers are replaced by finite similarity classes of terms,
will be investigated in \cite{Mannucci2}).
\\
\\
For the time being, let us imagine a concrete entity (an individual,
a finite physical machine, \ldots), named Amanda, trying to embrace
the natural number series. We shall imagine that we have some way to
gauge Amanda's level of confidence that a particular number is
within her reach. This measure will be coded by a certain  function
$G$ from the natural numbers into the $[0 \ \ 1]$ interval. For a
particular number $n$,  $G(n)=1$ shall indicate that $n$ is
definitely within reach, or \emph{strongly feasible}. On the
opposite side of the spectrum, $G(n)=0$ means that $n$ is
unfeasible, or totally beyond her grasp. All other values in between
will be considered feasible, in some  weak sense.
\\
\\
We shall make the assumption that the graspable number series (i.e.
the subclass of natural numbers that are presently not entirely
outside Amanda's reach), is closed downward and weakly closed under
some finite collection of elementary primitive recursive function
$F=\{ f_0, f_1, \ldots f_n \}$. By weakly closed we mean the
following: consider the algebra of closed terms that can be obtained
from $F$ from $0$. If $t_1, \ldots, t_n$ are terms in the algebra
and $G(t_1)= \ldots = G(t_n) =1$ and $f \in F$ is a n-ary function,
then $G(f(t_1, \dots, t_n))> 0$. The motivation behind the foregoing
is that it ought not to be possible to discriminate between the
strongly feasible and the unfeasible using any of the functions in
$F$.\footnote{A guiding mental image is the following: think of a
"castle" inhabited by the strongly feasible numbers. A "moat" of
weakly feasible numbers separates the castle from the vast world of
the unfeasible  beyond. The postulate above basically says that by
shooting "darts" from within the castle one never hits the
unfeasible: they all fall into the muddy moat in between. }
\\
\\
Let us start small: Amanda knows only $S$, the successor function
(or, to be more accurate, she knows only how to iterate $S$ a number
of time, depending on her actual resources. After a while, she will
probably fall asleep, or, in case she is a machine, run out of RAM).
So, the family F is $F=\{f_0=S\}$. The arithmetic world of Amanda is
encapsulated by the following definition:

\begin{defi}
A Fuzzy Initial Segment (FIS for short) of the Natural Number Series
is a  function  $G: N \longrightarrow [0 \ \  \ \ 1]$  such that
\\
\\
$1)$ \ \ $G(0) =1$  \ ( $0$ is a feasible number!)
\\
\\
$2)$ \ \ $\forall n \ \  G(n+1) \leq G(n)$   \  ( G is monotonically
decreasing)
\\
\\
$3$)\ \  $\forall n \ G(n)= 1 \Rightarrow G(n+1) >0$ \ (no jumps
allowed)
\\
\\
A FIS is \emph{strict} iff: \\
\\
$4)$ \ \  $\exists k $  $G(k) = 0$ \ (not every number is feasible)
\end{defi}
People even scantily aware of fuzzy logic will have recognized a FIS
as a special fuzzy subset of $N$. The intuition behind the foregoing
definition is that a FIS is downward closed as well as weakly closed
with respect to basic counting $0, S0, SS0, SSS0, \dots$.
\\
\\
The definition we just gave is a bit too general, at least for most
purposes: it includes some pathological fuzzy initial segments where
the rate of decrease in feasibility may temporarily slow down.
\footnote{ We are not ruling out a priori the possibility of
\emph{non-monotonic feasibility}. Quite to the contrary:  it is
perfectly conceivable that within certain contexts the rate at which
feasibility goes down as numbers get bigger will decrease, or even
that larger numbers may be considered more feasible (perhaps their
description has low complexity whereas some smaller number hasn't.
There is here an intriguing connection between Kolmogorov complexity
and feasibility, yet to be explored). Here, we are simply following
the old common sense precept: easier things first. } We can thus
restrict ourselves to more well behaved FIS:
\begin{defi}
A regular FIS (RFIS) is a FIS such that the function $ R(n) = G(n) -
G(n+ 1)$ is not decreasing, in the feasible segment $\{ n | G(n)
> 0\}$.
\end{defi}
A brief note on the last definition: the function $R(n)$ is an
important gauge for a FIS: the \textbf{one-step feasibility rate
change}. Our postulate $3)$ in Definition 1 rules out $R(n)= 1$, but
it is clear that a real-life FIS should have additional requirements
on its rate of feasibility loss. This topic will be elaborated
elsewhere, as it eventually leads to a general classification of
fuzzy initial segments of arithmetics. For the time being, though,
we shall stick to the above definition.
\\
\\
An elementary (but nevertheless interesting) example of a RFIS is
given by the the  family $ \{ G_N(n) \}$:
\\
\\
$(\clubsuit) \ \ G_N(n) = max(1 - \frac{n}{N}, 0)$
\\
\\
This family represent a set of fuzzy initial segments of arithmetic
where the degree of confidence decreases at the same pace
$\frac{1}{N}$, till the bottom is reached. If $N\uparrow \infty$,
then the FIS spans the entire natural numbers series; in other
words, \emph{all} finite natural numbers are feasible.\footnote{The
implicit suggestion here is that classical/intuitionistic reasoning
could be recovered by "passing to the limit": from the
ultrafinitistic standpoint this passage to the limit is obtained by
postulating an unbounded set of computational resources. }. Observe
that, according to our definition, only $0$ is strongly feasible in
$(\clubsuit)$. This is admittedly odd, but it can be easily fixed by
starting the decay at some positive number $n_0$.
\\
\\
Everybody knows that counting by "adding one" is extremely
inefficient: that is one of the underlying reasons for creating more
manageable notations. As in the mentioned "name the bigger number
game", the key is in devising fast growing primitive recursive
functions. Once such a function definition is achieved, it is
possible to reach out farther than before. Thus, a new "fuzzy
initial segment" is obtained. Going back to Amanda, let us assume
that a family $F= \{ f_0, \ldots, f_n \}$ of function is available
to her. Let us indicate with $\mathfrak{T}(F)$ the set of closed
terms generated by $F$. The arithmetic world of Amanda is now a
fuzzy initial segment of arithmetics weakly closed under $F$:
\begin{defi}
A Fuzzy Initial Segment closed under a family of functions $F$ is a
function $G: N \longrightarrow [0 \ 1]$ such that:
\\
\\
$1)$ \ \ $G(0) =1$  \
\\
\\
$2)$ \ \ $G$ is not increasing
\\
\\
$3$)\ \  if $t_1, \ldots , t_k$ are in $\mathfrak{T}(F)$  and  $f
\in \textit{A}$ is an n-ary function $\forall i \  G(t_i)=1
\Rightarrow f(t_1, \ldots , t_k)
> 0$
 \\
 \\
$4$) \ \ $\exists t \in  \mathfrak{T}(F)$ such that $G(t) = 0$.
\\
\\
Given two FIS,  $G$ and $G'$,  $G'$ dominates $G$  if $\forall k \
G(k) < G'(k)$.
\\
\\
The  number $n$  such that $\exists t \in \mathfrak{T}(F) \ (n=
length(t) \wedge G(t) > 0)$  and   $\forall t' \in \mathfrak{T}(F)
\neg \ (  G(t) > G(t') > 0  \wedge  length(t') < length(t) )$,
 is called the feasibility radius of the FIS.
\end{defi}
The \textbf{feasibility radius} is a gauge of how far out one can go
using the given a notation system.
\\
\\
Notice that a FIS in the sense of the previous definition is just a
particular case of Definition $2$, where $F=\{S\}$. The feasibility
radius is just the maximal $n=SS\ldots S0$ such that $G(n)>0$.
\\
\\
As we have already said, any FIS models a concrete arithmetic world
for a particular entity at a given time. Worlds change, and so do
arithmetic worlds. Amanda can expand hers in at least three ways:
\begin{enumerate}
  \item either by extending her feasibility radius (perhaps through memory
enhancers, or yoga \ldots), or
  \item  by being more creative and elaborating new ways of going farther with the same resources.
  \item Finally, she can move from her current $G$ to a $G'$ that has
exactly the same radius but such that $G'$ dominates $G$ (in other
words, her grasp on the feasible segment has gotten sharper).
\end{enumerate}
Examples of $1$ are easy to imagine, and we leave them to the
reader. As for $2$, let us go back to $(\clubsuit)$: it is clear
that the feasibility radius of $G_N(n)$ is $N-1$. Now, suppose
Amanda realizes that the unary successor notation is quite clumsy
(indeed, sticking to it, most of us would find the number $10000$
unfeasible!). Let us say that she discovers the binary digit
notation. Now, define a new FIS as
\\
\\
$G'(n) = G(\llcorner\log_2(n)\lrcorner + 1)$
\begin{prop}
Let G be weakly  closed under $f(m, n) = m + n + 2$. Then G' above
is closed under multiplication. Also, G' dominates G.
\end{prop}
The trivial proof uses the elementary properties of the logarithm:
$G'(m * n) = G(\llcorner\log_2(mn)\lrcorner)\geq
G(\llcorner\log_2(m)\lrcorner + \llcorner\log_2(n) + 2)$. Now, use
3) in Definition 3.
\\
\\
In the last example, Amanda's grasp has increased exponentially,
even though her memory/storage resources are the same (this trivial
example shows the power hidden in notation systems: through them we
can "handle" otherwise un-graspable  numbers).
\\
\\
As for $3$, compare $G_{2_N}(n)$ with $G_{2_N}^{\sharp} (n)= 1 -
\frac{2_n}{2_N}$: their radius with respect to successor is the same
( huge, even for small $N$!), but the shape of the FIS has changed
quite a bit. In $G_{2_N}^{\sharp}(n)$, things start out slowly
(making the first numbers almost strongly feasible), and
precipitates later on. Although the two FIS have the same length,
$G_{2_N}^{\sharp}$ clearly dominates $G_{2_N}$.
\\
\\
Before moving on, we wish to mention \emph{en passant} a
classification of feasible numbers into "size" (small, medium and
large), due to Kolmogorov (and taken up by Sazonov). Assume $G$ is a
FIS weakly closed under multiplication and such that $G(2) = 1$.
Define another FIS by $G^s(n) = G(2^n)$ (the superscript stands for
small)
\begin{prop}
The FIS $G^s(n)$ is closed under addition.
\end{prop}
The fuzzy set $S=\{ n \ | \ G^s(n) >0 \}$ is the collection of
$G$-small numbers. Further classifications can be obtained by
refining this method.
\\
\\
Though the arithmetic models we have presented are by no means the
most general, they are still broad enough to encompass not only the
"apeiron" view of UF, but also the "murios". Indeed, a model of
arithmetics with a largest number, such as the one proposed by
Mychielski \cite{Mychielski}( see for instance the interesting
article by Van Bendegem \cite{VanBendegem}), can be obtained by
"moding out" with respect to the following equivalence relation:
\\
\\
$n \sim m \Leftrightarrow (G(n)>0 \wedge G(m)>0 \wedge n= m) \vee
(G(n)= 0 \wedge  G(m)= 0)$
\\
\\
In plain words, for Amanda every number beyond the horizon is the
same number, aka infinity. Actually, there is yet another
equivalence relation producing a finite arithmetics model, namely
the one that identifies all numbers that are not strictly feasible.
Indeed, any "defuzzification operator" amongst the ones investigated
by Fuzzy Logic would do. By \emph{defuzzifying a FIS one ends up
with a crisp finite arithmetical universe}, which describes a
concrete arithmetical world where computational resources are
sharply known.
\\
\\
To summarize what we have seen, we can say that the  basic
arithmetical world may be enriched in a number of ways:
\begin{itemize}
  \item by considering \emph{a relativized form of feasibility}: as the
reader has certainly observed, the numbers that are feasible are the
ones that can be reached from $0$ in a small number of steps by
"jumping forward" with the aid of suitably devised terms.
\item by studying definable or even computable/effective Fuzzy Initial Segments, i.e. by putting
some restriction on the function $G$ (the reader has surely noticed
that so far no such restriction has been made).
\item
by considering a family of FIS, parameterized  by a time parameter,
representing different feasibility contexts in the life of Amanda
\item by introducing several "actors" (say, Amanda's boyfriend),
each equipped with his/her own FIS. This scenario appears to have
some similarities with multi-agents epistemic logics, but we are
unable to establish precise connections.

\end{itemize}
We shall not pursue these directions here, although we feel that
each one deserves further analysis. Instead, we will show that these
simple models are "enough" to provide an adequate semantics for
feasibly consistent arithmetics theories. But first a preliminary
analysis of proof theory from the perspective of UF is in order.

\section{ Dissipative Proof Theory}
In this section we briefly sketch a "revision" of standard proof
theory, which better accommodates the needs of  UF. Ordinary proofs
can be unfeasible, in the eyes of an ultrafinitistic mathematician:
their length, or other complexity measures he/she may want to
consider, could potentially turn them into proofs that would be
qualified as "ideal", and therefore unacceptable. It is then clear
that we need a way to gauge \emph{the degree of convincingness of a
proof}. At the same time, this should be done in the broadest
generality, so as to encompass different viewpoints.
\\
\\
Where shall we start? A proof is a tree built out of basic logical
rules. Each rule enables us to transfer the credibility of the
premisses into the consequences: if I believe that $A$ is the case
and I also believe $B$ is also the case, then, via
$\wedge$-introduction, I believe that $A \wedge V$ is true as
well.\footnote{For a view of natural deduction as a system of
assertions, the basic reference is  Martin L\"{o}f's  analysis in
\cite{MartinLof} } In standard proof theory, I can apply a rule as
many times as I like, or I can choose a different proof altogether:
if the hypothesis is held valid, the final conclusion will also be
considered valid, with the \emph{same degree of credibility}.
\\
\\
We intend to replace the foregoing implicit assumption  by a new
proof theory that allows for \emph{dissipative deductions}, i.e.
deduction where some degree of convincingness may be lost at each
steps. Just like classical formal arithmetics should be seen as a
limit case of feasible arithmetics, standard proof theory should be
viewed as a special case of \textbf{Dissipative Proof Theory}, where
the rate of truth loss is always zero, or negligible.\footnote{The
word dissipative is suggested by the following metaphor: think of
truth as some kind of "heat", flowing through the derivation, from
the leaves (i.e. the premisses). Depending on the medium through
which
heat flows, there can be some loss due to  dissipation.  } \\
\\
What we are suggesting here, is that each logical rule is both a
"joint", through which subproofs can be melded, and a
\emph{credibility transfer operator}, i.e. a channel that transforms
the credibility of the subproofs into the credibility of their
merger.
\\
\\
Concretely, we need a \textbf{Truth Transfer Policy}, i.e. a policy
that prescribes for each logical rule the degree of credibility of
the consequences a a function of the credibility of the premisses.
\\
\\
In the following we shall work within a Natural Deduction's
framework: $ND-Derivations$ will be the set of all correct
derivation from logical and non logical axioms. Everything can be
rephrased, \emph{mutatis mutandis}, in other systems of proof
theory. Also, as a "credibility thermometer", we shall choose the
unit interval $I=[0 \ \ 1]$, following Zadeh's fuzzy logic. It goes
without saying, though, that quite different measures of credibility
could be adopted, as long as a corresponding policy is also
introduced.
\begin{defi}

A Truth Transfer Policy (TTP)   $\mathfrak{P}$,  is an
assignment,for each logical rule $R$, of a function
$\mathfrak{P}_R:I^k \longrightarrow I$, where $k$ is the number of
premisses in the rule. $\mathfrak{P}$ will obey the following
restrictions:
\begin{itemize}

  \item The TTP of the axioms is the identity $id: I \rightarrow I$.
   \item $\mathfrak{P}_R (0, \dots , 0) = 0$ for each rule.

  \item $\mathfrak{P}_R (1, \dots , 1) > 0$ for each rule.

  \item $\mathfrak{P}_R (p_1, \dots , p_k) \leq min (p_1, \ldots, p_k) $

\end{itemize}
Given a TTP $\mathfrak{P}$ and a derivation $p$, we can associate an
$I$-value to it, its $\mathfrak{P}$-credibility, by assigning to all
the logical and non logical axioms a credibility value of $1$, and
applying $P$ at every steps of the deduction. This recursive
procedure produces a function
\\
\\
$F_\mathfrak{P}: NP-Derivations \longrightarrow [ 0 \ \ 1]$ \\
\\
that will be referred to as  the $\mathfrak{P}$-based
\emph{credibility measure}.
\\
\\
We shall say that $A$ is a\emph{ feasible consequence} of the theory
$T$ under the TTP  $\mathfrak{P}$ iff there is a proof $p$ of $A$
from $T$ such that $F_{\mathfrak{P}}(p)>0$.  If
$F_{\mathfrak{P}}(p)=1$, then $A$ is a \emph{strong consequence} of
$T$.
\\
\\
We shall use the notation $T\vdash_{F_{ \mathfrak{P}}} A$ to say
that $A$ is feasibly derivable from $T$ under $F$. If the derivation
is strong, we shall indicate it as $T\vdash_{F_{ \mathfrak{P}}}^{!}
A$
\\
\\
A theory $T$ is $\mathfrak{P}$-feasibly consistent under $F$ there
are no feasible proofs under $\mathfrak{P}$ of $\perp$ from $T$.
\end{defi}
Let us briefly see what consequences Definition $5$ entails: we have
assumed that $\mathfrak{P}$ has no credibility jumps, in order to
make sure that
\\
\\
1) $T\vdash_{F_{ \mathfrak{P}}}^{!} A$ and $T\vdash_{F_{
\mathfrak{P}}}^{!} B $ implies $T\vdash_{F_{ \mathfrak{P}}} A\wedge
B$
\\
\\
2)$T\vdash_{F_{ \mathfrak{P}}}^{!} A$ and $T\vdash_{F_{
\mathfrak{P}}}^{!} A \Rightarrow B $ implies $T\vdash_{F_{
\mathfrak{P}}} B$
\\
\\
and similarly for the other connectives. In general, it will not be
true that $T\vdash_{F_{ \mathfrak{P}}} A$ and $T\vdash_{F_{
\mathfrak{P}}}A \Rightarrow B $ implies $T\vdash_{F_{ \mathfrak{P}}}
B$: merging already barely feasible proofs of $A$ and $B$ may go
beyond the horizon, constructing an unfeasible proof.
\\
\\
The reader may be a bit puzzled: what is the connection of the above
with the usual way of thinking about feasibility of proofs, namely
their length? Our TTP encompasses this view as a very special case;
indeed, looking at the length of a proof as a measure of its
convincingness basically corresponds to assigning the \emph{same
constant erosion factor }to each logical rule. In other words, the
only impact that a rule application has on the credibility of a
proof in progress is simply the one that results from the increase
of size of the proof itself.
\\
\\
To illustrate the foregoing with a rather trivial example, let us
assume, for sake of simplicity, that instead of Natural Deduction we
operate with an Hilbert-style system. Also, in order to simplify
matters, let us imagine that only Modus Ponens has a non trivial
dissipative policy. More specifically,  if I have a proof of $A$ of
credibility $p_A$ and I have a proof of $A\Rightarrow B$ of
credibility $p_(A\Rightarrow B)$, then a single-step application of
modus ponens will lead to a proof of $B$ of credibility
\\
\\
$ \mathfrak{P}(MP) =   f: [0 \ \ 1]^2  \longrightarrow [0 \ \ 1]$
\\
\\
where $f =\max ( 0, \min( p_{(A\Rightarrow B)}, p_A ) - E)$
\\
\\
and $ 0< E < 1$ is the "constant credibility erosion factor".
\\
\\
Now, suppose that we have a basic arithmetical theory, say $Q$, with
a feasibility predicate, $F()$, obeying
\\
\\
$F(0), F(n)\Rightarrow F(n +1)$, but $\neg F(2^{1000})$ (no
induction on the predicate $F$ !).
\\
\\
It is immediate to check that posing $E= \frac{1}{2^{1000}}$, makes
the theory feasibly consistent: the convincingness (obtained by
iterated modus ponens) of $F(n)$ is only $1- \frac{n}{2^{1000}}$. As
trivial as this example is, it already reveals a couple of
interesting things:
\begin{itemize}
  \item first, that for small numbers the credibility is almost equal one,
creating, as it were, the illusion of an indefinite series of
feasible numbers. Indeed, standard Hilbert style proof theory
corresponds to setting the constant erosion factor equal zero;
  \item secondly, it implicitly suggests where a FIS itself may come from:
it is simply generated by the proof-theoretical discoveries of
Amanda. Posing $G(n) =1- \frac{n}{2^{1000}}$ we recover the trivial
FIS described in $(\clubsuit)$ of the previous Section.
\end{itemize}
The point of view presented in the foregoing could be qualified as
\emph{local}: the overall credibility of a proof is the result of
"integrating" over all the one-step credibility changes caused by
applying such and such rule in the course of the proof.
\\
\\
It is quite possible that \emph{non-local} measures of credibility
might become relevant as well. For instance, Carbone and Semmes have
advocated a measure of complexity of proofs based on the number of
cycles in the logical flow graph of the proof itself (see
\cite{Carbone}). In turn, one could take their measure, compose it
with a notion of feasibility for natural numbers, and obtain a
credibility measure of proofs. Such a measure is non local, at least
at first sight; it is not clear to us whether it could ever be
recovered as a TTP-grounded measure through some kind of roundabout
procedure, but it seems quite unlikely at this stage. At any event,
this order of considerations implicitly suggests the following
definition:
\begin{defi}
A general credibility measure (GCM)  for derivations is simply a
fuzzy characteristic function  $F: ND-Derivations \longrightarrow [0
\ \ 1]$
\\
\\
A GCM  $F$ is \emph{well-behaved} iff  $T\vdash_F A \
\Leftrightarrow \  T, \neg A $ is $F$-consistent. \footnote{We omit
for brevity the definition of feasible consequnce, feasible
consistency, etc. They are just the same as in the case of a GCM
that is generated by a TTP (see page 20).}
\\
\\
A GCM $F$ is \emph{factorable} iff there exist a complexity measure
\\
\\
$c:ND-Derivations \longrightarrow N$
\\
\\
and a FIS
\\
\\
$G: N \longrightarrow [0 \ \ 1]$
\\
\\
such that $F= G \circ c$:
\\
\\
\\
 
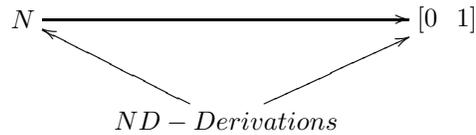
\xymatrix{ N\ar[rr] &   & [0 \ \ 1] \\
 &   ND-Derivations \ar[lu] \ar[ru] & }
 \\
 \\
\end{defi}
In his proposal for an ultrafinitistic proof theory (see again
\cite{Sazonov}), Sazonov introduces two restrictions to ordinary
proof theory: proofs must be physically realizable, and normal (or,
which is equivalent, cut-free if one uses Gentzen's approach). To
us, the first prescription needs to be \emph{formalized}, as it is
not absolute (feasibility is contextual!). This can be easily
accomplished by choosing the GCM obtained from a distinguished FIS
(representing here the semi-set of feasible, i.e constructible
numbers on a concrete machine), and using as the complexity measure
the standard 'number of symbols". As for his second restriction,
namely normality of proofs, it can be absorbed by the following
observation: suppose one has a GCM $F:
\\
\\
ND-Derivations \longrightarrow [0 \ \ 1]$
\\
\\
and a computable map
\\
\\
$T: ND-Derivations \longrightarrow : ND-Derivations$
\\
\\
one can define a new GCM by composition
\\
\\
\\
 
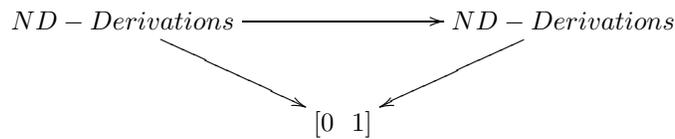
\xymatrix{ND-Derivations \ar[rr] \ar[rd] &   & ND-Derivations \ar[ld]\\
 &  [0 \ \ 1] & }
\\
\\
\\
Indeed, the schema above seems to underly  Sazonov's choice. Why
banishing a rule in the first place? A sensible answer is that (as
it has been pointed out in several quarters) the cut rule "hides"
the actual complexity of a derivation. By "unwrapping" it, one has a
better sense of how complicated the proof actually is. The
unwrapping is accomplished precisely by setting $T$ above as the
standard cut-elimination procedure.
\\
\\
We have just barely introduced the bare bones of Dissipative Proof
Theory. We just wish to remark that the notion of $TTP$ forces a
paradigm shift even in the very way one thinks of proofs:
derivations, from being \emph{barren trees}, become
\emph{decorated}. \footnote{ A note in passing for the
categorical-minded reader: this suggests that the topic of
Dissipative Proof Theory could possibly benefit from an injection of
ideas and method from Enriched Category Theory}.
\begin{que}

 Which factorable credibility measures are of the form
$F_\mathfrak{P}$ for a suitable TTP  $\mathfrak{P}$ ?
\end{que}
\begin{que}
The notion of well-behaved CGM appears to be critical (see next
Section). It would thus be important to know which TTP are such that
their generated GCM are well-behaved. Also, which measures of
complexity and FIS combine to form well-behaved GCM?

\end{que}
What we have seen in this Section has to do with credibility, not
Aristotelian-Tarskian truth. Nevertheless, it is clear that to
re-establish a comfortable parallel between provability and truth,
between what-can-be-proved and what-is-out-there, a careful revision
of the core foundations of Model Theory is necessary. This is  our
goal in the next Section.

\section{Vague Truth: a semantics \`{a} la Tarski for feasibly consistent  theories}
Before we begin, let us pause a moment and point out that the very
distinction between syntax and semantics, the cornerstone of
classical logic, is suspect from an ultrafinitistic viewpoint. To be
sure, even from a more conservative constructivist perspective,
people have raised a number of objections against the dualism
syntax-semantics (Jean-Yves Girard has recently written on this
pivotal issue in \cite{Girard}). All these objections remain true
\emph{a fortiori} for ultrafinitism (a Tarski-free semantics for
arithmetics inspired by formal games will be described in
\cite{Mannucci2}).
\\
\\
So, why even bother trying to establish a tarskian semantics for
feasibly consistent theories? AS we declared in the introduction,
our chief goal is to provide a bridge between the classical world
mathematics and the world-to-be of ultrafinitism. As Tarskian
semantics, grounded in Set Theory, is pervasive, it seems reasonable
to start from there. Moreover, we would like to turn the informal
and intuitive models of Section 4 into formal ones, thus providing a
way to talk about feasibly consistent theories without being
hampered by the "bureaucracy of syntax" ( we are borrowing here the
colorful expression coined by Girard).
\\
\\
The Tarskian notion of satisfaction, upon which the entire cathedral
of standard model theory hinges, is unfit to provide a semantics for
UF. Indeed, as it stands, it cannot even accommodate traditional
constructivism, and we are here in a territory far more "picky" than
all brands of liberal constructivism (i.e. constructivism that
allows for potential infinity in its arguments) can possibly be.
\\
\\
We thus need to revise the "rules" of satisfaction, and replace the
abstract notion of truth underpinning the Tarskian definition with a
more realistic one. \\
\\
The key is in the TTP introduced in the last Section:\footnote{In
the following, we shall restrict ourselves to TTP-generated
deduction systems, as those system give us "control" on the flow of
truth. Also, for sake of simplicity, we shall confine ourselves to a
subset of ND which includes rules only for the set $\{\wedge,\ \neg,
\  \exists  \} $ (classically, this set is complete).   To which
extent our arguments can b lifted to more general context is not
clear at present. } the TTP for a specific connective should act as
an \emph{lower bound }for the "degree of truth" in a given model.
Assume in a model $\mathcal{M} $ (yet to be defined) $A$ holds with
truth degree $p_A$, and $B$ with degree $p_B$ (where $p_A, p_B \in
[0 \ \ 1])$. The degree of truth of $A \wedge B$ should be at least
the degree of truth of its proof from the premisses $A$ and $B$
using $\wedge$-introduction. In other words, we are not ruling out
the possibility that in $\mathcal{M} $ additional credibility may be
available for $A \wedge B$ through some kind of evidence; what we
are saying is that its degree of truth should be at least the one
inferred using a one-step proof from $A$ and $B$.

\begin{defi}
Let $\mathcal{L}$ be a first-order language and  $\mathfrak{P}$ an
assigned Truth Transfer Policy. A structure for $(\mathcal{L},
\mathfrak{P})$ is an n-tuple $\mathcal{M}= ({D}, {c_k}
,{f_i},{P_j})$ where domain, constants and function symbols are
interpreted the usual way, whereas  each $n$-ary predicate $P$
becomes a fuzzy subset of $(D)^n$: $ \textbf{P}: (D)^n
\longrightarrow [0 \ \ 1]$
\\
\\
Given an assignment $s$ of the variables, we say that $\mathcal{M}$
satisfies the atomic predicate $P(x_1, \ldots , x_n)$ iff
$\textbf{P}(s(x_1),\ldots, s(x_n))>0$. In symbols, $(\mathcal{M}, s)
\models P$. If the evaluation is $1$, then we say that $M$ strongly
satisfies $P$ ($(\mathcal{M}, s) \models_{!} P$).
\\
\\
Satisfaction is extended to arbitrary predicates by posing
constraints to the shape of their corresponding fuzzy subsets (we
shall present satisfaction rules only for the complete set
$\{\wedge,\  \neg, \  \exists  \} $   ):
\\
\\
1)\  $\textbf{P}(x_1, \ldots , x_n)\ \wedge \ \textbf{Q}(y_1, \ldots
, y_m)\geq \mathfrak{T}_{\wedge-intro}(\textbf{P}(x_1, \ldots ,
x_n),\textbf{Q}(y_1, \ldots , y_m)) $
\\
\\
2) $\neg\textbf{P}(x_1, \ldots , x_n) = 1- \textbf{P}(x_1, \ldots ,
x_n)$
\\
\\
3)\ $\exists x \  \textbf{P}(x)\geq
\mathfrak{T}_{\exists-intro}(\textbf{P}(c)) $, where $c$ is any
constant in $\mathcal{M}$.
\\
\\
A $\mathfrak{T}$-model $\mathcal{M}$ for a theory $T$ is a structure
such that all axioms of  $T$  are strongly true in $\mathcal{M}$.

\end{defi}
A comment to the above is in order: the notion of satisfaction in
the foregoing definition is (as the careful reader has certainly not
failed to observe) strictly coupled with a specific $TTP$. This may
be puzzling at first, but it suffices to see that standard
first-order semantics does exactly the same thing, only in disguise.
Indeed, Tarski's truth is simply a particular case of the previous
definition in which the underlying $TTP$ is assumed to be the
trivial one, i.e zero-decay: \\
\\
\emph{within classical logic truth is preserved in its entirety at
every logical step}.
\\
\\
Definition 6 gives "legal status" to FIS as models of theories.
Recall once again the theory $T_0=Q + \neg F(2^{1000})$ of the
previous section. Consider the structure $\mathcal{M_0}= (N, =, S,
+,*, G)$, where $G$ is the FIS $G(n)= 1- \frac{n}{2^{1000}}$ (the
interpretation of equality is the usual one). $M_0$ is a model of
$T_0$, according to Definition 6. Similarly, one can construct
models for other arithmetical theories enriched with a feasibility
predicate.
\\
\\
A good semantics ought to be at the very least sound. This is indeed
the case:
\begin{prop} Soundness.
Let $T$ be a first order theory. If $T\vdash_{\mathfrak{T}} A$, then
for every $\mathfrak{T}$-model $\mathcal{M}$ of $T$,
$\mathcal{M}\models A$

\end{prop}
The proof is, just like its standard counterpart,  by induction on
the complexity of $A$.  $\Box$
\\
\\
Notice that something more can be extracted by the last proposition:
the inequality holds for \emph{each} proof of $A$ from $T$.
\begin{cor}
$A(x_1, \ldots, x_n) \geq \min \{ F_{\mathfrak{T}}( p_A) \}$, where
$p_A$ denotes a proof of $A$ from $T$.
\end{cor}
The last inequality implicitly suggests that somewhere there must be
a model where equality holds, i.e. a model where the degree of truth
is exactly what one gets by choosing, as it were, the "best" proof
available. This is indeed the case, when one consider the familiar
construction of the term model, a key ingredient in the Henkin's
version of the completeness theorem for FOL. We are  ready to
establish an ultrafinitistic version of the completeness theorem.

\begin{prop} Completeness. Let $\mathfrak{T}$ be a TTP such that the corresponding GMC
is well-behaved. If for every $\mathfrak{T}$-model $\mathcal{M}$ of
$T$ $\mathcal{M} \models A$, then $T\vdash_{\mathfrak{T}} A$.

\end{prop}
Sketch of proof: if it is not true that $T\vdash_{\mathfrak{T}} A$,
then the theory $ T= T \bigcup \{\neg A \}$ is
$F(\mathfrak{T})$-feasibly consistent (here the well-behavior of
$\mathfrak{T}$ is needed). The goal is to show that this theory has
a model, thereby contradicting the assumption.
\\
Just like in the standard completeness theorem, we can  construct a
special model, the term model $\mathcal{M}_0$, of a feasibly
consistent completion of the theory $T + \neg A$. The key is
defining truth in $\mathcal{M}_0$: the value of a given predicate
$A^M(x_1, \ldots, x_n)$ will be the minimum credibility value over
all  $F(\mathfrak{T})$-feasible proofs of $A^M(x_1, \ldots, x_n)$
from $T$. If no such proofs exists, one can "feasibly complete"
$\mathcal{M}_0$ by adding either $A^M(x_1, \ldots, x_n)$ or its
negation to it (unless, of course, $T$ feasibly proves the negation,
in which case there is no need to expand $T$). In other words, the
construction mimics the familiar construction of the term model, but
where the role of consistent completion is taken by feasible
consistency. Indeed, Henkin's proof should be recoverable when one
"passes to the classical limit". $\Box$
\\
\\
Before we conclude this Section, we would like to remark that the
notion of feasible consistency is akin to the so-called
para-consistent theories, studied by several authors
(\cite{Carnielli}). Indeed, in our Definition 6, a structure can
satisfy a formula and its negation at the same time. However, in our
setting, it is not the logic that has changed, as in para-consistent
systems. It is the very notion of deducibility and truth that has
been "upgraded", to account for more stringent requirements than
either classical logic or intuitionistic one can provide. It would
still be useful to see if a bridge between these two approaches can
be drawn. Much remains to be done here \ldots

\section{Cantorian Nanotechnology: Miniaturizing \\Cantor's Paradise}
We have just seen that it is possible to "miniaturize" the notion of
natural number series and, through a suitable emendation of syntax
and semantics, view these objects as genuine \emph{nano-models} of
standard arithmetical theories. In a metaphorical way, this is a bit
like Nanotechnology, only applied to Logic instead of Engineering:
choose a macroscopic object, and strive to create a microscopic
clone, all the while retaining its salient properties.
\\
\\
In the same spirit, one can think of the (bounded) finite as a
microscopic world, and the transfinite (or even the unbounded
finite) as the macroscopic one. The challenge then seems to be: to
what extent can we miniaturize familiar infinitary structures? For
instance, can we create an entire theory of transfinite cardinals
once we have a finitary copy of $\Omega_0$ via a FIS? These
cardinals would form a ladder of transcendency degrees of
finiteness, just as the alephs gauge the size of standard infinite
sets.
\\
\\
This program should by no means  remain confined to structures.
Instead, it should attempt to miniaturize basic mathematical and
meta-mathematical results as well. For instance, as we have seen in
the last Section, a coherent ultrafinitistic version of the
celebrated completeness theorem can be  formulated (possibly
different ones: we are not making any claim that our account is the
only possible route). What about the incompleteness theorems? Or the
Lowenheim-Skolem theorem? Armed with a notion of satisfaction such
as the one we have just described, one could start the exploration
of concrete first-order mathematical theories (such as the theory of
infinite dimensional vector spaces) that do not afford conventional
finite models.
\\
\\
In our belief,  this program, which we would like to refer to as
\textbf{Cantorian Nanotechnology}, is a viable channel of
investigation. At the very least, it will give us a better sense on
the real boundaries (assuming there are any) between the finite and
the infinite. If successful, it could provide us with a cornucopia
of new entities that may have an impact in the way we model our
physical world \footnote{ A single example: quantum mechanics
represents a physical system as a Hilbert space. Many quantum
systems, such as the free spinless particle on the line, are
associated to an infinite dimensional Hilbert space. In BN, the same
particle could be modeled by a subspace $H$ of $\mathcal{C}^N$,
where $N$ is a large enough finite integer; $H$ would appear as
infinite dimensional. Incidentally, we observe \emph{en passant}
that such a novel perspective could be used to motivate approaches
to Quantum Mechanics from the Quantum Computing's angle. }and our
very selves.
\\
\\
Last, but by no means least, Cantorian Nanotechnology could be an
attempt to move beyond the lasting debate between constructivists of
all brands and staunch defenders of Cantor's Paradise: it may turn
out, after all,  that the beautiful land that Cantor created for the
endless joy of mathematicians, can be faithfully reproduces  at the
finite scale.
\\
\\
\\
\textbf{Aknowledgments}
\\
\\
The authors would like to thank our common friend Juliette Kennedy
for kindly inviting one of us to present a talk based on this paper
at the Tennenbaum Memorial on April 6th 2006, held at The Graduate
School of CUNY under the umbrella of MAMLS (Mid-Atlantic
Mathematical Logic Seminar). Thanks go also to the other MAMLS
organizers, David Joel Hamkins, Roman Kossak and Arthur Apter.
\\
\\
Two brief, white-heat discussions with Domenico Napoletani (while
sipping his fabulous teas after a strenuous practice of standing
qigong), were greatly beneficial in clarifying the overarching goals
of our project. Some of his interesting comments will find their way
in \cite{Mannucci1}.
\\
\\
In recent months, the first author has posted a few messages on the
FOM list (see for instance \cite{FOM1}, \cite{FOM2}), concerning
ultrafinitism and related issues. Several FOM subscribers
(including,  but not limited to, Vladimir Sazonov, David Isles,
Stephen Simpson, and Karlis Podnieks) replied, publicly or
privately,  with valuable comments, criticisms, suggestions, and
different viewpoints. To them all goes our gratitude.
\\
\\
Last and foremost, our eternal gratitude goes to the late Stanley
Tennenbaum: his iconoclastic attitude and insistence on keeping a
child's eyes even in the midst of the most abstruse and seemingly
difficult topics, was, and is, the true driving force behind this
project.

\noindent Mirco A. Mannucci\\
Department of Computer Sciences \\
George Mason University \\
e-mail: mmannucc@cs.gmu.edu
\\
HoloMathics, LLC\\
e-mail: mirco@holomathics.com
\\
\\
\noindent Rose M. Cherubin\\
Department of Philosophy,\\
George Mason University\\
e-mail: rcherubi@gmu.edu


\begin{thebibliography}{99}

\bibitem{BigNumbers} Scott Aronson, {\bf Who Can Name the Bigger Number?}, available at
http://www.scottaaronson.com/writings/bignumbers.html
\bibitem{AndreevGordon} P. V Andreev and E.I Gordon, {\bf A Theory of
Hyperfinite Sets}, available at
http://www.math.huji.ac.il/~logic/external/ths.pdf
\bibitem{Hajek}Petr Hajek,
{\bf Metamathematics of Fuzzy Logic}  Kluwer, 1998
\bibitem{HajekPudlak} P. Hajek and P. Pudl\'{a}k, {\bf Metamathematics of First-Order Arithmetic}. Springer-Verlag, 1993.
\bibitem{Nelson}Ed Nelson {\bf Faith and Mathematics } available at http://www.math.princeton.edu/~nelson/papers/s.pdf
\bibitem{Nelson2} Ed Nelson,  {\bf Predicative Arithmetics}, available at http://www.math.princeton.edu/~nelson/books/pa.pdf
\bibitem{Parikh} Rohit Parikh {\bf Existence and Feasibility in
Arithmetics} JSL 36, 1971
\bibitem{VanDantzig} Van Dantzig, {\bf is $10^{10^{10}}$ a natural
number?} Dialectica, vol 19, 1955
\bibitem{Vopenka} Petr Vopenka, {\bf Mathematics in the Alternative
Set Theory}, Teubner, 1979
\bibitem{Sazonov}Vladimir Sazononv, {\bf On feasible numbers}, available at
 http://www.csc.liv.ac.uk/~sazonov/papers.html
\bibitem{Isles1} David Isles, {\bf On the notion of standard non-isomorphic natural number
series} n Constructive Mathematics, F. Richman (ed.), Springer
Verlag, pp.111-134, 1981
\bibitem{Isles2}  David Isles, {\bf Finite Analog to the Loewenheim-Skolem
Theorem} Studia Logica, Vol 53, pp. 503-532, 1994
\bibitem{Volpin}A. S. Yessenin-Volpin, {\bf The ultra-intuitionistic criticism and
the antitraditional program for foundations of mathematics}, in
Intuitionism and Proof Theory, A. Kino, J. Myhill, and R. E. Vesley,
eds., North-Holland, 1970, pp. 1--45.
\bibitem{Troelstra} Anne S. Troelstra and Dirk Van Dalen, {\bf Constructivism in Mathematics}, Vol. 1 \& 2. North-Holland 1988.
\bibitem{Lavine} Shaughan Lavine {\bf Understanding the Infinite},
 Harvard University Press, second edition 1998.
\bibitem{Jouko} Jouko Vaananen  {\bf A Short Course on Finite Model Theory}, available at
http://www.math.helsinki.fi/logic/people/jouko.vaananen/jvaaftp.html
\bibitem{Cherubin} Rose M. Cherubin, Mirco A. Mannucci
{\bf Muryon and Apeiron: The
 very large and the indefinite in early greek thought and mathematics}, in
preparation.
\bibitem{FOM1} Mirco A. Mannucci, {\bf Feasible and Utterable
Numbers}, FOM posting, at
http://cs.nyu.edu/pipermail/fom/2006-July/010657.html

\bibitem{FOM2} Mirco A. Mannucci, {\bf Feasible Consistency-Truth Transfer Policy}, FOM posting, at
http://www.cs.nyu.edu/pipermail/fom/2006-October/011016.html

\bibitem{Mannucci3} Mirco A. Mannucci   {\bf Models of Ultrafinitism II:
A Roadmap for  Feasible Realizability}, in preparation.

 \bibitem{Mannucci2} Mirco A. Mannucci {\bf Models of Ultrafinitism III:
Deconstructing the Term Model}, in preparation.

\bibitem{Mannucci1} Mirco A. Mannucci {\bf Models of Ultrafinitism IV:
A pragmatic
 approach to Cantor's Theory of Cardinalities}, in preparation.
\bibitem{Beck1} Jon M. Beck {\bf Simplicial sets and the foundations of analysis}.
 In Proceedings of Conference on Sheaf Theory,Durham, England(July 1977),
 volume 753 of Lecture Notes in mathematics, pages 113-124. Springer, Berlin, 1979.
\bibitem{Beck2}Jon M. Beck. {\bf On the relationship between algebra and analysis}.
Journal of Pure and Applied Algebra, 19:43-60, 1980.
\bibitem{Gerla}Giangiacomo Gerla,
{\bf Fuzzy Mathematical Tools for Approximate Reasoning.} Kluwer
Academic Publishers, Dordrecht, 2001. xii+ 266
 pp.

\bibitem{MartinLof} Per Martin L\"{o}f, {\bf On the meanings of the logical constants and the
 justifications of the logical laws}. Nordic Journal of Philosophical Logic 1 (1): 11-60.
 Lecture notes to a short course at Università degli Studi di Siena, April 1983.

\bibitem{Girard} Jean-Yves Girard {\bf Les fondements des
mathématiques}, 2003, available at
http://iml.univ-mrs.fr/~girard/Articles.html


\bibitem{VanBendegem}Jean Paul Van Bendegem {\bf Classical
Arithmetics is quite unnatural} Logic and Logical Philosophy, Volume
II, 2003.
\bibitem{Mychielski} Jan Mychielski, {\bf Analysis without Actual
Infinity}, Journal of Symbolic Logic, vol. 46, number 3, 1981.
\bibitem{Carbone} A.Carbone, S. Semmes, {\bf Looking from the inside and from the outside}, Synthèse, 125: 385-416, 2000.


\bibitem{Troelstra2} Anne Siep Toelstra, Scwichtenberg, {\bf Basic
Proof Theory}, Second Edition, Cambridge Tracts in Theoretical
Computer Science, CUP, 2000
\bibitem{Borel} Emil Borel {\bf Les Nombres Inaccesibles} avec une note de M. Daniel DUGUÉ. Paris, Gauthier-Villars, 1952.
\bibitem{Archimedes} Archimedes, {\bf The Sand Reckoner}, available
at http://www.math.uwaterloo.ca/navigation/ideas/reckoner.shtml
\bibitem{Kahn} Charles H. Kahn,  {\bf Anaximander and the Origins of Greek
Cosmology},   Columbia University Press, 1960; corrected edition,
Centrum Press, 1985; reprint Hackett, 1994.

\bibitem{McKir}  Richard D. McKirahan, Jr.,  {\bf Philosophy Before Socrates}, Hackett, 1994.

\bibitem{Pecus} Rose Cherubin, {\bf Why Matter? Aristotle, the Eleatics, and the Possibility of Explanation},
Graduate Faculty Philosophy Journal, vol. 26, no. 2, pp. 1-29, 2005.

\bibitem{Pecus2} Rose Cherubin, {\bf Inquiry and What Is: Eleatics and Monisms},
Epoché vol. 8, no. 1, pp. 1-26, 2003.

\bibitem{Ari} Aristotle, {\bf Physica},  edited by W.D. Ross. Oxford Classical Texts. Oxford: Oxford University Press, 1950.

\bibitem{Simplicius} Simplicius,  {\bf In Aristotelis Physicorum. Commentaria in Aristotelem Graeca IX}, edited by Hermann Diels. Berlin: Reimer,
1882.
\bibitem{Carnielli} Walter Carnielli, {\bf How to build your own
paraconsitent logic}, available at
ftp://logica.cle.unicamp.br/pub/e-prints/Carnielli.pdf

\end{thebibliography}
\end{document}